# Modelling the Structure and Dynamics of Science Using Books

Michael Ginda
Cyberinfrastructure for Network Science Center (CNS)
ILS, SOIC, Indiana University, USA
mginda@indiana.edu

Andrea Scharnhorst
KNAW, The Netherlands
andrea.scharnhorst@dans.knaw.nl

Katy Börner
Department of Information and Library Science, CNS, SOIC
Indiana University, USA
katy@indiana.edu

## Introduction

Scientific research is a major driving force in a knowledge-based economy. Income, health and well-being depend on scientific progress. The better we understand the inner workings of the scientific enterprise, the better we can prompt, manage, steer, and utilize scientific progress. Diverse indicators and approaches exist to evaluate and monitor research activities—from calculating the reputation of a researcher, institution, or country to analyzing and visualizing global brain circulation. However, there are very few predictive models of science that are used by key decision makers in academia, industry, or government interested to improve the quality and impact of scholarly efforts.

Other scientific communities rely extensively on predictive models to simulate events such as weather, seismic hazards (UNAVCO Facility, 2010), or epidemics (Colizza et al., 2006). Recent efforts have sought to forecast science and technology in the form of an "innovation accelerator" (Van Harmelen et al., 2012). However, the heterogeneous and partially commercial datasets required to model science remain scattered; cultures of algorithm and model sharing are slow to evolve; and a unified theory that interlinks validated models of science does not yet exist.

According to the Oxford English Dictionary (2002), the term model may function as: a representation of structure or system; an object of imitation; and a type and design. The latter two definitions of model are used to indicate an objects status as an exemplar meant to be imitated or a prototype to be copied and are irrelevant for what is discussed in this chapter. The first function, i.e., a representative model, is the focus here and may either describe a targeted system or phenomena (e.g., a science model); represent a broader theoretical interpretation of the laws, axioms, and models of a discipline (e.g., a model of science); or perform both functions simultaneously (Frigg & Stephan, 2012).

In this chapter—building on prior work (Scharnhorst et al., 2012)—we define a model of science as "a systematic description of an object or phenomenon that shares important characteristics with its real-world counterpart and supports its detailed investigation" (Börner et al., 2012a, p 1). Models of science put forward a theoretical and/or empirical understanding with predictive power and they are validated based on the accuracy of their predictions. Focusing on scientific models of science, we purposefully exclude anecdotal evidence and narratives, e.g., the analysis of science fiction literature to identify possible future developments (Steinmueller, 2010). Instead, we focus on models of science that explain and help predict the activities of scholars (also called authors, researchers, scientists) because they are the generators of ideas and innovation—papers don't write papers, authors do (Cronin, 2005)—and it is scholars that collaborate and read and write papers leading to the diffusion of ideas, knowledge, and innovations and the "making of science" (Cronin, 2008).

The remainder of the chapter is organized as follows. The next section discusses challenges and opportunities when attempting to delineate and map the space of existing models of science. Subsequently, we present a novel "bibliographic-bibliometric" analysis which we apply to a large collection of books relevant for the modelling of science—we explain the data collection together with the results of the data analyses and visualizations. In the final section we discuss how the analysis of books that describe different modelling approaches can inform the design of new models of science.

## Prior Work: Context and Focus

Models of science are developed in many scientific disciplines. They use different (mathematical) approaches and terminology that is hard if not impossible to align across disciplinary boundaries.

Descriptive models of science can be found in the field of philosophy of science, history of science, sociology of science, and science and technology studies, in short in all those areas of social sciences and humanities which have knowledge production as their object of study. Bernal's encyclopedic work "The Social Function of Science" (1939, 1967) has influenced many of those reflecting about science in a scientific manner (Garfield, 2007). Since 1981, the Society for Social Studies of Science[1] awards the *John Desmond Bernal*

---

[1]  http://4sonline.org/

*Prize* annually to scholars that have made a distinguished contribution to the field. The first three award recipients were Derek de Solla Price (1981), Robert K. Merton (1973), and Thomas S. Kuhn and their books *Little Science, Big Science* (Price, 1963), *The Sociology of Science* (Merton, 1973), and *The Structure of Scientific Revolutions* (Kuhn, 1962) are included in this analysis.

Predictive models of science (computational and mathematical) are developed in scientometrics, bibliometrics, system dynamics, physics and mathematics, but also more recently in a new branch of philosophy of science and cognition (Payette, 2012). One of the first predictive models was introduced by Goffman—he used a model originally developed to predict the spread of diseases to describe the spreading of ideas (Goffman & Nevill 1964; Goffman 1966; Harmon 2008). The so-called SIR model orders researchers in three categories: the number researchers 'susceptible' to a new idea but not yet infected with it (S), the number of 'infected' researchers (I), and the number of 'recovered' researchers (R) who lost interest and will not return to the idea. The model presumes that boundaries of scientific fields and/or invisible colleges (Crane, 1972) can be defined. Goffman's work showcases the complex relationship between mathematical, theoretical models and their empirical validation. Using his model, it is possible to define with which possibility a researcher becomes 'infected' with an idea and the predicted growth rate of a new scientific field can be compared with the actual growth rate (Wagner-Döbler, 1999), see review of follow-up studies in (Lucio-Arias & Scharnhorst, 2012)). Case studies also show that it is not easy to validate all processes inscribed in Goffman's model (Burger & Bujdoso, 1985).

Goffman's model is only one out of many approaches to conceptualize science. As of today, there exists no unifying framework that would interlink models—neither in terms of co-author communities nor in terms of citation linkages between publications. When (Lucio-Arias & Scharnhorst, 2012) presented a bibliometric study of a set of relevant LIS journals and the perception of the three scholars Lotka (1926), Price (1965; 1976), and Goffman (1966), they found that "Mathematical models of the sciences are divided into different branches and exist largely in isolation" (p.35).

There are very few comparisons of existing models or attempts to combine multiple models to arrive at a more holistic understanding of the structure and dynamics of science. Textbooks that provide an overview of different types of models can only be written if an acknowledged and shared body of validated models exists—which is not yet the case, though an inventory of models in certain domains has been attempted, e.g., see Scharnhorst et al. (2012) and Schulze (2014).

As with any system, there are many different ways one can study and model the science system. Some scholars study science by looking into its cognitive and logical structure (Collins, 1988), or into its political-economic base (Nowotny et al., 2005), its institutions, its politics, its social actors (Gibbons et al. 1994), or its communications (Kaufer & Carley 1993). Even those cognizant to the problems of studying science are scattered across philosophy of science, history of science, science and technology studies (STS) and scientometrics. Among them, philosophy and history of science have their own scientific societies, journals and conferences and only occasionally meet. Concerning STS, as represented in the *Society for Social Studies of Science* or the *European Association for the Study of Science and Technology*[2], one can observe groups of researchers that perform qualitative studies exclusively but also groups that perform only quantitative studies. There are few who try to bridge between these two different epistemic perspectives, and even fewer who reflect about science in a wider historic context of knowledge production. Among them, Blaise Cronin stands out as a scholar able to play on all strings of the harp of scientific reflection about science. He looks at current forms of scholarly communication from a view point which encompasses scholarship from the Enlightenment to Force11 (Cronin & Sugimoto, 2014). His early book *The Citation Process* (1984) called for a study of science as a social system taking into account "norms and values which guide and constrain the actions of individual scientists" (p. 1).

In this chapter, we investigate the location and interlinkages of books that are relevant for the development of models of science. World Cat data[3] of library catalog records and subject headings plus library classification codes were used to identify a set of relevant books, to identify major topical clusters, and to show interlinkages. The resulting semantic networks were then explored to determine the spheres of influence, relevance, and context around specific sets of books on models of science, subject headings, and library classification codes assigned by librarians around the globe.

### Bibliographic-Bibliometric Data Collection and Analysis

Currently there exists neither a "Models of Science" handbook nor a comprehensive annotated bibliography. A search for "models of science", "models of science dynamics", "modeling processes of science", "modeling of scholarly communication", or similar using the Web of Science, Scopus, Google Scholar is of limited value when aiming to identify relevant literature. Our starting point is the collection *Models of Science Dynamics* (Scharnhorst et al., 2012) that presents a review of major types of and applications for models of science. While this book does not claim to cover all relevant works across the landscape of science, the authors of each chapter reviewed a specific branch of models of science developed in different areas of science. Using references to books on modelling science, library classification data and subject headings can be retrieved and used to map the evolving topical space in which models of science are researched and developed.

---

[2] http://easst.net/
[3] World Cat is a database managed by the Online Computer Library Center (OCLC) that collects library catalog records from around the world into a single information resource discovery system. http://www.worldcat.org

### Identification of Relevant Books

To map the concept "model of science," a book list was generated using the references from the *Models of Science Dynamics* book (Scharnhorst et al., 2012). Using a bibtex file that captured all 589 references cited in the book, 196 citations were identified as book references. Two additional books were added: the *Models of Science Dynamics* (2012) book itself and the book *The Web of Knowledge: A Festschrift in Honor of Eugene Garfield* (2000). The latter was edited by Blaise Cronin and Helen Barsky Atkins. It was added to this analysis as a landmarks review and outlook in the field of science studies but also in honor of this Festschrift.

*Identification of Associated World Cat Subject Headings*

The resulting list of 198 books was then searched in World Cat to collect all English language subject terms and to determine the accuracy of the document type. Twenty-one titles were removed from the seed list for three reasons: (a) the citation was not a book [e.g., conference proceedings that were not published as a book, and therefore not cataloged (9), journal articles (3), or self-published program instruction manuals (4)]; (b) the book reference lacked subject headings in English (4); and (c) the book reference duplicated a book in the data (1). For the frequency distribution of the final 177 titles by type, *book, ebook, incollection*, and *inproceedings*[4], see Table 1.

**Table 1:** Number of reference types of book titles

| Initial Reference Types | | Final Reference Types | |
|---|---|---|---|
| article | 3 | book | 147 |
| book | 151 | ebook | 1 |

---

[4] Each citation collected for this analysis had a bibtex category assigned that defined its genre. The categories include: *article* indicates that a citation is an article published in a *book* and *ebook* that indicate that a citation is either a book or an electronic book without a print publication; *electronic (handbook)* indicates that a citation is for software tool handbooks; *inproceedings* indicates that a citation was published in a conference proceeding rather than a book; last, the category *incollection* indicates that a citation is chapter included a multi-author or edited book.

| | | | |
|---|---|---|---|
| ebook | 1 | incollection | 18 |
| electronic (handbook) | 1 | inproceedings | 11 |
| incollection | 21 | | |
| inproceedings | 21 | | |
| **Grand Total** | **198** | | **177** |

For a distribution of all 198 book titles and the 177 final books references per publication year (binned by 5-years) see Figure 1. Most of the cited books in *Models of Science Dynamics* are published between 2001 and 2005 (bin label 2005). This age-distribution for cited work is in line with other studies on obsolescence of literature (Lariviére et al., 2008) but could also signal the relative youth of this domain.

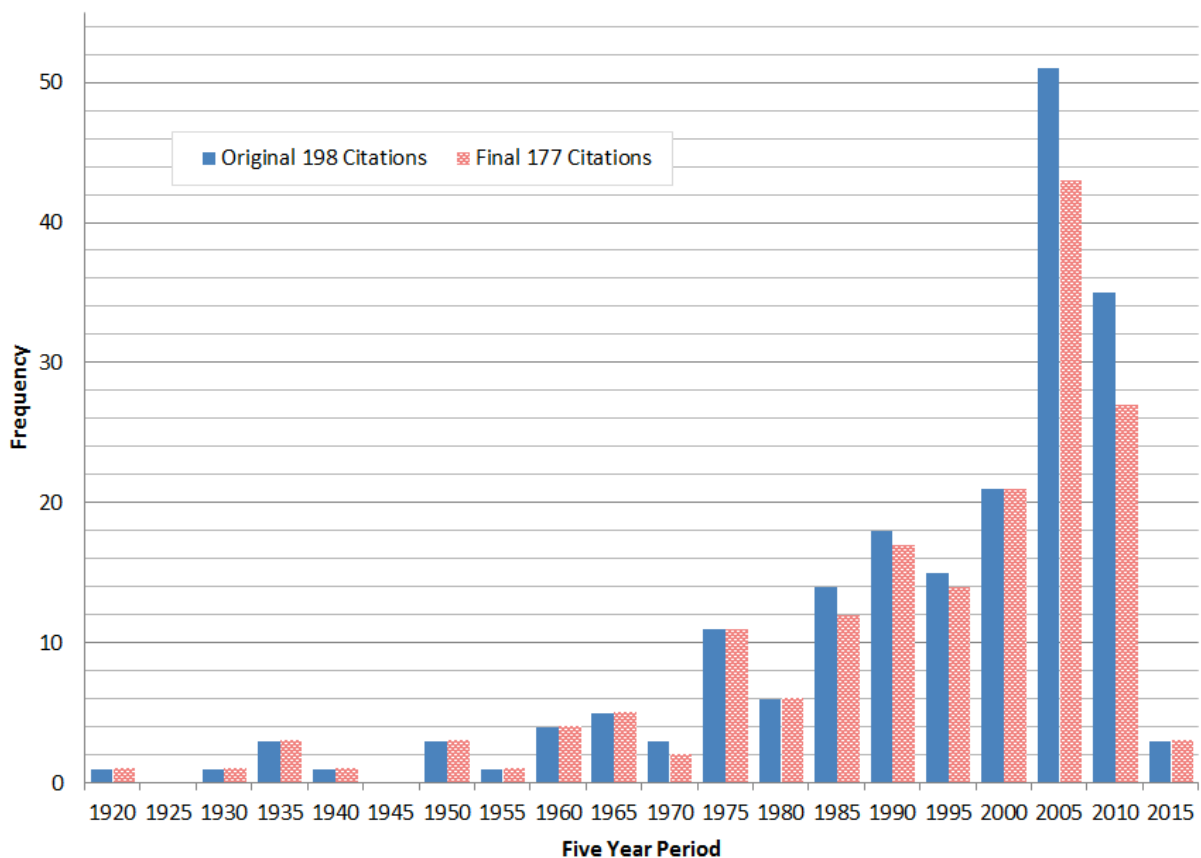

**Figure 1:** Number of initial 198 (blue) and final 177 book titles (red) per publication year.

The bibliographic record for books in World Cat contains among others a field called *subject.* This field contains the *subject headings, genre terms and forms, and unindexed subject keywords* attributed to a book by a librarian or cataloguer when the book is purchased and added to the collection of a library. Modern information systems may allow librarians to look up already attributed subject headings for a work, and wide-spread bibliographic classification systems as Dewey, Unified Decimal Classification or Library of Congress lead to some standardization. Still, libraries have individual classification schemes and different indexing practices and librarians around the globe may assign rather different subject headings to the same book.

To harvest the various subject headings, we selected "View all editions and formats" in publically displayed bibliographic records of a book in World Cat. Collecting from all unique editions of a book allowed us to gather the full variety and scope of subjects assigned by catalogers around the globe. This method provided a substantial number of different subject headings for each book and all distinct terms per book were identified using a semi-automatic process.

In a second step, the collected subject headings needed to be normalized for spelling, topicality, and relevance. The initial list of 1,313 subject headings for all 177 books was consolidated into a list of 876 unique subject headings after removing duplicate occurrences. The unique subject headings were refined a second time to combine related topics and remove extraneous headings. Subject headings were *combined* if: the subject varied in spelling or punctuation (e.g. *Biology – Mathematical models*[5] also includes *Biology / Mathematical model*); the subject heading contained a designation of the type of material (e.g., *Biology – Mathematical models – Textbooks* would appear under *Biology – Mathematical models*), geographic region (e.g. *Alcoholism and crime – Wales – Cardiff* is grouped with *Alcoholism and crime*), or temporally (e.g., *Economic history – 16th century* would appear under *Economic history*); the subject headings is topically similar enough that a work could be found using the chosen variant (e.g., *Biophysics/Biomedical Physics* is grouped under *Biophysics*; *Comprehension (Theory of knowledge)* is grouped under *Comprehension*). Subject headings were *removed* if they described the materiality of a book (e.g., electronic book), a geographic place without a proceeding topic (e.g., Japan, Great Britain), or were the name of a researcher (e.g., Lotka). The final list contains 675 unique subject headings.

These subject headings are distributed unevenly over the 177 books. Books usually carry more than one subject heading. On the average, there were 6.31 subject headings per book, ranging from 1 to 31. A book's set of subject headings indicate the topics that indexers and catalogers determine are coextensive to the work, i.e., the concepts that most accurately represent a book's subject. Coextensive subject headings indicate a co-occurrence relationship between concepts. The co-occurrence of subject heading have been used to identify inter-index consistency and to map concept-spaces based on indexer perceptions of subject headings (Olson & Wolfram, 2008; Gabel & Smiraglia, 2009). Within this analysis, the co-occurrence of subject headings across multiple books is used as a proxy measure of the relationships between science domains, see analysis of this data in the next section.

---

[5] Throughout the text, book subject heading are *italicized;* subject heading domains groups are **bold.**

## Identification of Associated Library of Congress Classification Codes

Next, we examined how the 177 books were classified. Using the Library of Congress online catalog,[6] the Library of Congress Classification (LCC) shelf numbers[7] were collected for 171 books (six books did not have LCC shelve number). The number of books and the number of subject headings for each of the nine LCC classes is given in Table 2. For example, seven of the 171 books have been classified under **B - Philosophy, Psychology, Religion**. How these seven books and their subject headings distribute over the next level in the classification is also shown in the table. That is, the table interlinks LCC classes to books, and subject headings. Note that different subject headings might appear simultaneously in different LCC classes.

**Table 2:** Library of Congress Classifications and Respective Book and Subject Heading Counts

| Library of Congress Classification | Book Count | Subject Heading Count |
|---|---:|---:|
| **B - Philosophy, Psychology, Religion** | **7** | **36** |
|     B - Philosophy - General | 2 | 12 |
|     BF - Psychology | 2 | 12 |
|     BC - Logic | 1 | 7 |
|     BD - Speculative philosophy | 1 | 4 |
|     BJ - Ethics | 1 | 1 |
| **H - Social Science** | **63** | **429** |
|     HM - Sociology | 18 | 155 |
|     HB - Economic Theory, Demography | 15 | 75 |
|     H - Social Sciences - General | 9 | 52 |
|     HD - Industries, Land use, Labor | 8 | 68 |
|     HC - Economic history and conditions | 4 | 30 |
|     HV - Sociology - Social pathology… | 3 | 21 |
|     HQ - Sociology - The family… | 2 | 8 |
|     HA - Statistics | 2 | 6 |
|     HF - Commerce | 1 | 9 |
|     HG - Finance | 1 | 5 |
| **J - Political Science** | **1** | **10** |
|     JN - Political Institutions… | 1 | 10 |
| **L - Education** | **2** | **25** |
|     LC - Special Aspects of Education | 2 | 25 |
| **Q - Science** | **77** | **563** |
|     Q - Science - General | 38 | 253 |

---

[6]    Library of Congress Online Catalog http://catalog.loc.gov/vwebv/searchAdvanced
[7]    Library of Congress Classification codes http://www.loc.gov/catdir/cpso/lcc.html

| | | |
|---|---|---|
| QH - Natural History, biology | 15 | 141 |
| QA - Mathematics | 13 | 98 |
| QC - Physics | 7 | 50 |
| QP - Physiology | 2 | 13 |
| QD - Chemistry | 1 | 4 |
| QL - Zoology | 1 | 4 |
| **R - Medicine** | **2** | **23** |
| RC - Internal Medicine | 1 | 12 |
| RA - Public Aspects of medicine | 1 | 11 |
| **T - Technology** | **10** | **121** |
| T - Technology - General | 5 | 42 |
| TK - Electrical Engineering… | 4 | 65 |
| TA - Engineering - Civil Engineering | 1 | 14 |
| **U - Military Science** | **1** | **8** |
| UG - Military Engineering, Air forces | 1 | 8 |
| **Z - Bibliography. Library Science…** | **8** | **63** |
| Z - Books (General), Writing… | 7 | 56 |
| ZA - Information resources | 1 | 7 |
| **Books without LCC Codes** | **6** | **27** |
| Total | 177 | 1305 |

Next, we use book LCC numbers to define a crosswalk of LCC classes to a wider scientific domain coding system, see Table 3. We use the thirteen major scientific disciplines identified in the UCSD Map of Science (Börner, et. al., 2012b) as a proxy for upper-level knowledge organization. For example, **QH, QP, QL** are assigned to **Biology**. Four domains from the UCSD map did not appear in the LCC codes: **Health Professionals**, **Infectious Diseases**, **Biotechnology**, and **Earth Sciences**. **Earth Sciences** was given a code because it could not be subsumed under a secondary code; **Biotechnology** is grouped with **Biology**, **Health Professionals** is grouped with **Medicine**, and infectious diseases is grouped with **Epidemiology** in the **Social Sciences**. A **Science General** category was also added to categorize subjects that either could be applied across domains (e.g. the subject heading Research is coded zero) or relates a specific domain's study of science broadly (e.g., **Science – social aspects** is coded zero and twelve to indicate connection between social science and the general study of science).

**Table 3:** LCC class and science domain code crosswalk, with related book counts

| Code | Domain | LCC Class | Book Count | Notes |
|---|---|---|---|---|
| 0 | Science General | Q | 37 | Subjects that can be applied across domains. |
| 1 | Biology | QH, QP, QL | 18 | UCSD domain Biotechnology grouped here. |
| 2 | Medical Specialties | R (all) | 2 | UCSD domains Health Professionals grouped here. |

| 3 | Engineering | T, TA, UG | 5 | LCC class T is split between code 3 and 6. |
|---|---|---|---|---|
| 4 | Chemistry | QD | 1 | |
| 5 | Earth Science | - | 0 | |
| 6 | Electrical Engineering & Computer Science | T, TK, Z, ZA | 14 | Library and Information Science included |
| 7 | Brain Research | BC | 1 | Cognitive Science and Psychology |
| 8 | Humanities | B, BD, BF, BJ, LC | 8 | History, Philosophy, Education |
| 9 | Math & Physics | QA, QC | 20 | |
| 10 | Social Sciences | H (all), JN | 65 | Sociology, Economics, Business, etc. UCSD Infectious Diseases grouped here. |

The division of domains within LCC classes does not align cleanly with the domains identified by the UCSD map. In particular, **Social Science** books are dispersed across and combined within LCC class divisions, while works related to modern interdisciplinary technology are classified within the general **Technology** class. This is not a surprise. Classification or knowledge organization systems have a history and moreover are tailored towards the collection for which they are designed (Smiraglia, 2014).

Later, we then applied the same domain coding system to assign the book subject headings a scientific domain using a common code book. Each subject heading was assigned one to two of the eleven domain codes shown in Table 3. A subject heading's domain code was identified by analyzing its topic coverage and the LCC number(s) assigned to the book(s) using the particular subject heading. The goal of coding subject headings in this manner is to see where topics (as expressed by subject headings broadly) overlap across domains and disciplines. We treat subject headings as terms of a controlled vocabulary. Individually or in combination with each other they characterize a topic. Domain codes were applied in an as needed fashion; some subject headings have only one associated domain; and secondary domain codes were only made when the subject is studied across multiple domains. Subject headings with broad application were mapped into the domain code zero. The resulting code matrix is not balanced because some subject headings were given both a primary code for a domain most associated with a subject; the secondary code indicates the second domain associated with a subject; likewise, many subjects were not coded twice.

The result of these multiple mappings is a co-occurrence of domains by the number of subject headings associated with both domains related to books on modeling science, see Table 4.

**Table 4:** Cross tabulation of domain codes assigned to book subject headings

| Domain Name | Domain Code | 0 | 1 | 2 | 3 | 4 | 6 | 7 | 8 | 9 | 10 | Single Domain Code | Grand Total |
|---|---|---|---|---|---|---|---|---|---|---|---|---|---|
| Science (General) | 0 | | 1 | | | | 1 | | 2 | 6 | 15 | 17 | 42 |
| Biology | 1 | 2 | | 3 | | 3 | | | 4 | 13 | 3 | 23 | 51 |
| Medicine | 2 | | 4 | | | | | | | 2 | 4 | 3 | 13 |
| Engineering | 3 | | | | | 1 | | | 2 | 5 | 6 | 5 | 19 |
| Chemistry | 4 | | 1 | | 2 | | | | | 3 | | 6 | 12 |
| Earth Science | 5 | | 3 | | | | | | | | 2 | 0 | 5 |
| Elect. Eng. & Comp. Science | 6 | 3 | | | 6 | | | | 1 | 23 | 29 | 56 | 118 |
| Brain Research | 7 | 1 | 7 | | | | 1 | | 5 | 2 | 24 | 9 | 49 |
| Humanities | 8 | 2 | 1 | | | | 3 | | | 1 | 15 | 10 | 32 |
| Math & Physics | 9 | 2 | 4 | 1 | 7 | 2 | 14 | | 1 | | 10 | 61 | 102 |
| Social Sciences | 10 | 5 | 5 | | | | 23 | 1 | 50 | 21 | | 127 | 232 |

Of the 675 unique subject headings, 317 subjects were coded with one domain. Conversely, 358 subject headings were assigned two domain codes as they were either complex subject headings, multiple concepts and domains imbedded in them (e.g., the subject heading *Science – Psychological aspects* would be coded for **Science (General)** and **Psychology**) or a subject heading topic was associated with multiple domains (e.g., the subject *Social Networks* is a methodology used within the **Electrical Engineering & Computer Science** and **Social Sciences** domains). The subject *Communication in science – Data processing* was coded **Science (General)** and **Electrical Engineering & Computer Science** because *Communication in science* could refer to research by any number of domains, while the secondary topic *Data processing* is a topic most relevant to **Information and Computer Science**[8].

The domains of **Social Sciences** (sociology, economics), **Math & Physics**, and **Computer and Information Science** are most strongly associated with subject headings from the 171 books, followed by **Biology, Psychology**, and general science domains. Domains with the most domain intersections are highlighted in red: **Social Sciences** and **Humanities** (50); **Electrical Engineering & Computer Science** and **Social Sciences** (29); **Brain Research** and **Social Sciences** (24); **Electrical Engineering & Computer Science** and **Math & Physics** (23); **Social Sciences** and **Electrical Engineering & Computer Science** (23); **Social Sciences** and **Math & Physics** (21). Please note that these intersections are created by the content of our specific set of books. In other words, books relevant to modelling science combine knowledge between social sciences, mathematics (and physics), computer and information science. This also entails that to be able to study models of science, readers and authors needs to be familiar with several areas of research.

---

[8] Throughout the text, book subject heading are *italicized;* subject heading domains groups are **bold.**

## Topical Space of Books Relevant for Modelling Science

Using the data detailed in the previous section, different topical spaces can be extracted, analyzed, visualized, and interpreted.

### Major Subject Headings Linked to Books

To understand the topical space of books and subject headings, a bipartite network of the 177 books and their 675 subject headings was extracted. The resulting network has 852 nodes—too many to depict in a network layout in letter size. Using the Science of Science Tool (Sci2)[9] and the Gephi[10] graph visualization platform, the network was analyzed to identify all subject nodes with an out-degree (i.e., number of linked books) greater than five, and all their associated books. The resulting network has 19 subject nodes and was laid out in a two dimensional space using a force-directed layout (Figure 2).

Subject heading nodes are colored pink and labeled by subject headings; the nodes for the 177 books are green and labeled with book titles. For the 19 subject heading nodes, node and label size increases and color darkens as the out-degree increases from six to 21. For book title nodes, the node and label size and the color are scaled according to the number of unique subject headings associated with a book title in the original network. Node labels are truncated to improve the readability of the graph.

---

[9] http://sci2.cns.iu.edu
[10] http://gephi.org

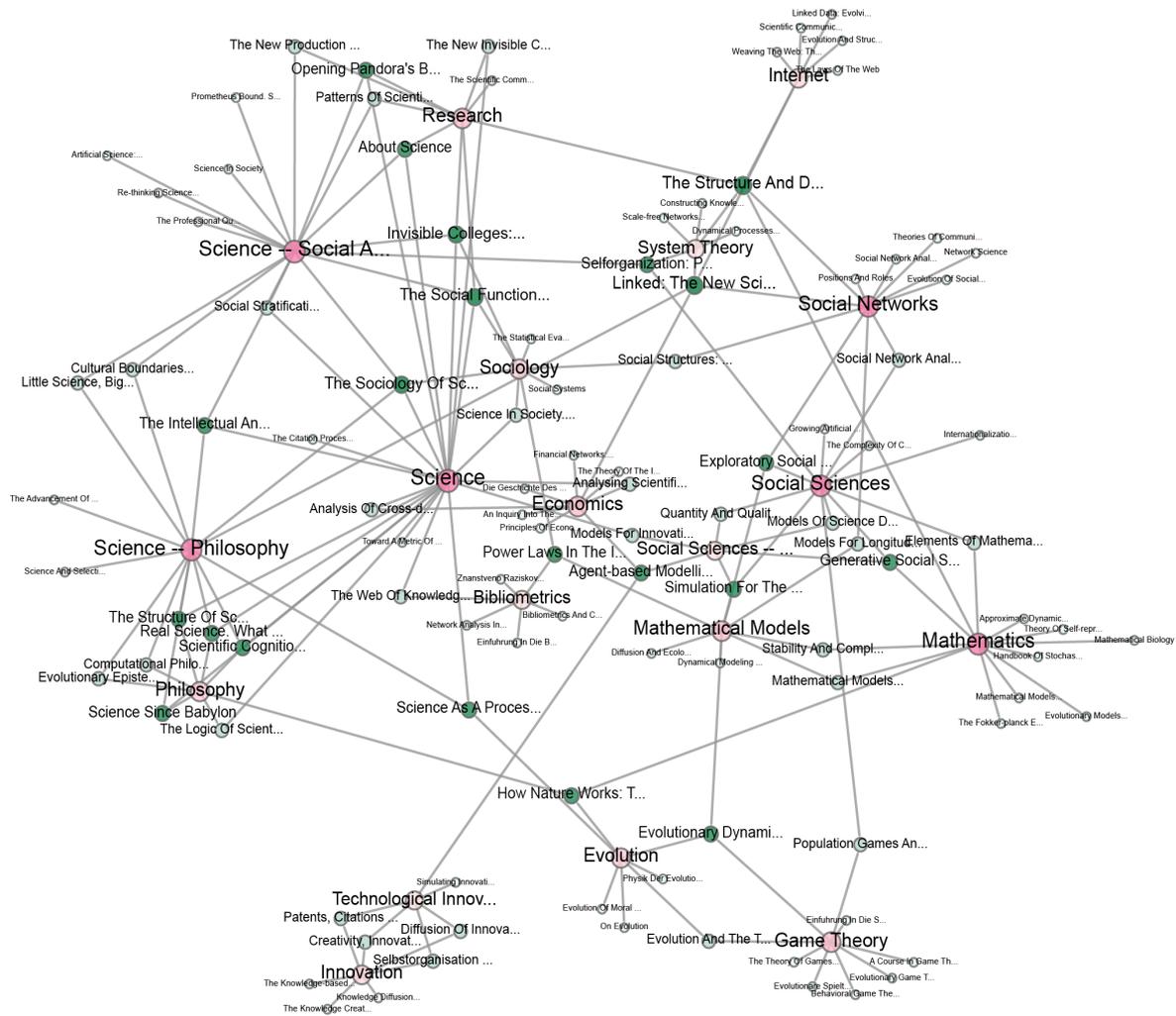

**Figure 2:** Bimodal network of high-degree subject heading nodes (in pink) linked to associated book nodes (in green). See website at http://cns.iu.edu/2015-ModSci.html for a high-resolution, searchable pdf file.

Overall, the network shows that the high-degree subject headings and associated books cover a wide range of modelling approaches developed in diverse disciplines of science. *Science* characterizes many of the books, and its subcategories S*cience – Social Aspects* and *Science – Philosophy* play a major role. We also see "Mathematics" and "Mathematical models". But beyond that, we see specific areas in mathematical modeling emerging: *System Theory*, *Game Theory*, models of *Evolution* and *Social Networks*. Another set of subject headings describes research areas that inspired new mathematical models of science to answer questions related to the process of knowledge production, e.g., *Innovation*, *Communication*, *Internet*, and *Bibliometrics*. While most books are associated with only one subject heading, some books are associated with many areas. Among them is *The Structure and Dynamics of Networks* (Newman et al., 2006) which introduces the highly interdisciplinary, emerging area of network science to a broad audience. Usually, books belonging to the same epistemic thread are connected to the same subject headings. For instance, Per Bak's book *How Nature Works* (1996) is among others linked to the subject node *Evolution* which contains other books that reflects about evolution either from a point of view of physics (physics of self-organization), or game theory (evolutionary game theory) or biology. One of them is the German title *Physik der Evolutionsprozesse* (Ebeling et al., 1990)—a linkage that would be difficult to identify using linguistic analysis or citation-based analysis. Those two books belong to one research stream within statistical physics. Some titles do not deal with science specifically, but describe methods that can be applied to describe and model complex phenomena such as the science system itself. Only a close inspection of the content of the books can reveal this similarity yet subject headings and library classification codes can be used to identify key linkages.

Figure 3 shows the same network using the very same node positions. However, the Blondel community detection[11] algorithm (Blondel et al., 2008) was applied randomly using a resolution parameter of 0.9. The networks modularity was measured to be 0.643. The communities detected in this network represent the major areas of research on models of science discussed in *Modelling Science Dynamics* (Scharnhorst et al., 2012), including, the philosophy of science and knowledge (teal), the science studies (yellow), innovation and communication in science (pink), economics and social sciences (purple), mathematical models (lime green), bibliometrics and information science (Kelly green), evolution and game theory (light blue), and computer science (blue).

The science studies community encompasses books from science and technology studies, as the *New Production of Knowledge* by Gibbons et al. (1994), the classics *Invisible Colleges* by Crane (1972) from the sociology of science, as well as *The New Invisible College* by Wagner (2008) from bibliometrics. Also "Science" as general subject heading is put into this community. Note that *The Web of Knowledge: A Festschrift in Honor of Eugene Garfield*, edited by Blaise Cronin and Helen Barsky Atkins (2000) (indicated by a red dotted frame) bridges two major communities relevant to study science: the community of science studies "Science" and the community of "Bibliometrics."

---

[11] The Blondel community detection algorithm partitions a network into communities based on the density of links in a network. A node's membership in a Blondel community is determined by its relationship to other nodes. Nodes are more likely to link to members within their community, than link to those outside of their communities. The algorithm detects and partitions communities based on the relative density of the relationship between nodes in a given network.

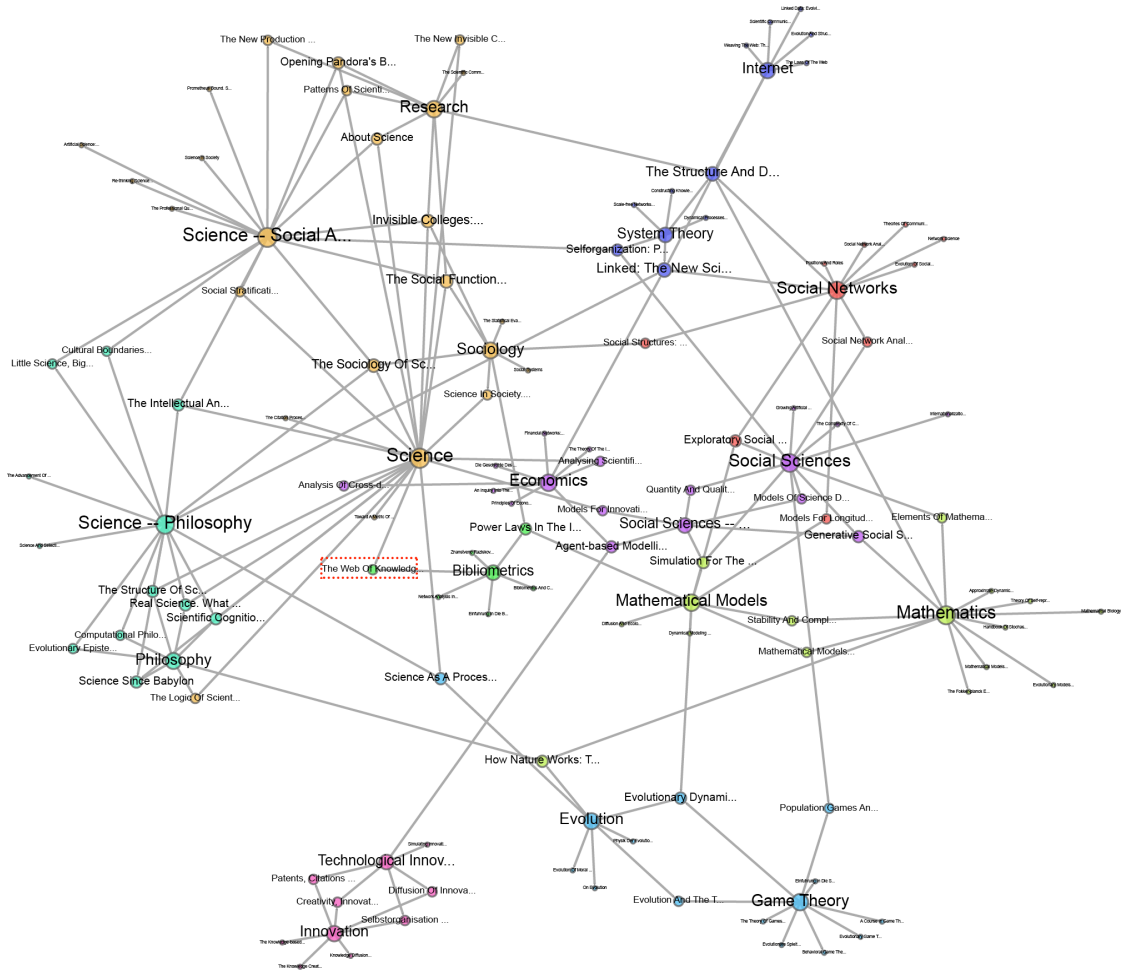

**Figure 3:** Bimodal network of high-degree subject heading nodes linked to associated book nodes and colored by Blondel communities. See website at http://cns.iu.edu/2015-ModSci.html for a high-resolution, searchable pdf file.

Just like in Figure 2, the different subject headings are grouped by *disciplines*, *methods*, and *perspectives*. The community detection algorithm and coloring groups different subject headings, e.g., *Science – Philosophy* and *Philosophy*. It also makes visible how aggregations of subject headings are interlinked, e.g., *Mathematical Models* and *Mathematics* are closely interlinked with *Social Sciences*.

Figure 4 shows how the topic space of models of science has developed over time. We use the publication date of the different books (first print) and color coded book nodes by binned years. Books published between 1750 and 1990 are given on the left-hand side. 18 of the 19 subject headings are shown—only *Internet* is missing. Several books are not yet published. The full network is given on the right hand side—with book nodes colored by year bins. Early books in yellow and later books in cyan and green can be easily identified.

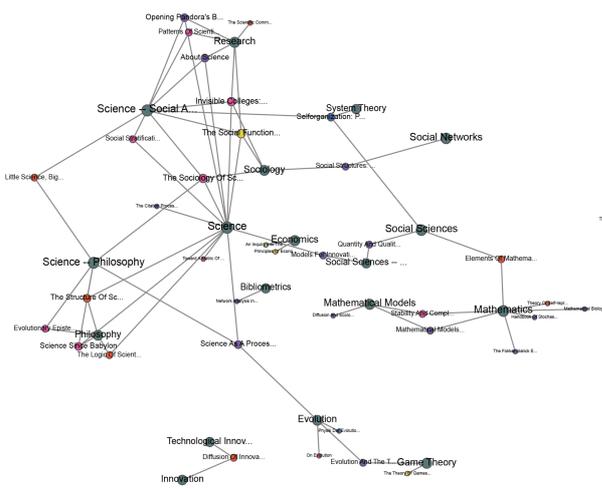 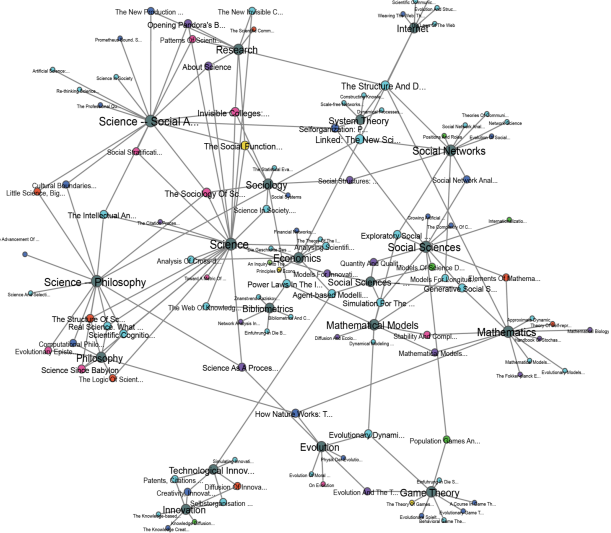

Books Published between 1750 and 1990    Books Published between 1750 and 2011

Legend

○ 1750-1900  ○ 1901-1949  ○ 1950-1969  ○ 1970-1979  ○ 1980-1989  ○ 1990-1999  ○ 2000-2010  ○ 2011  ○ Subject Headings

**Figure 4:** Temporal comparison of networks by book publication date for period between 1990 and 2011. See website at http://cns.iu.edu/2015-ModSci.html for a high-resolution, searchable pdf file.

## Major Books in the Topic Space

Let's have a closer look at the book *Models of Science Dynamics* (Scharnhorst et al. 2012). The book is divided into three major parts: The "Foundations" part covers two introductory chapters; the "Exemplary Model Types" part introduces three different types of models such as epidemics models, agent-based models, and game theoretic models; the "Exemplary Model Applications" part showcases the application of different models to study collaboration and citation networks. The chapters are written by author teams from different scientific disciplines and they cite different areas of work. Exactly 14 books are cited in more than one chapter and possibly in the Foreword (FW) or Preface (PF), see Table 5. Among the books listed in Table 3, Kuhn's *Structure of Scientific Revolution* (1962) stands out, followed by Price *Little Science, Big Science* (1963). While written in the 1960s, both still inspire today's modelling science efforts.

**Table 5:** Listing of books that are cited by more than one chapter, the Foreword, or Preface

| Title of the book (authors) | Year | Chapters[12] |
|---|---|---|
| *The Theory of Games and Economic Behaviour* (Von Neumann, Morgenstern) | 1944 | 2,5 |
| *Human Behaviour and the Principle of Least-Effort* (Zipf) | 1949 | 1,3 |
| *The Structure of Scientific Revolutions* (Kuhn) | 1962 | FW,PF,1,2,3,6 |
| *Little Science, Big Science and Beyond* (Price) | 1963 | FW,PF,1 |

---

[12] PF=Preface, FW=Foreword

|   |   | ,3,4,6 |
|---|---|---|
| *Invisible colleges: Diffusion of knowledge in scientific communities.* (Crane) | 1972 | FW,1,6 |
| *The sociology of science: Theoretical and empirical investigations* (Merton) | 1973 | PF,1,6 |
| *Matematicheskie modeli v issledovanii nauki* (Yablonsikii) | 1986 | PF,FW |
| *Introduction to informetrics: quantitative methods in library, documentation and information science* (Egghe & Rousseau) | 1990 | FW,2,3 |
| *The New Production of Knowledge. The Dynamics of Science and Research in Contemporary Societies.* (Gibbons et al.) | 1994 | 1,6 |
| *Social network analysis: Methods and applications* (Wasserman & Faust) | 1994 | 6,7 |
| *Growing artificial societies: social science from the bottom up* (Epstein&Axtell) | 1996 | 2,4 |
| *Linked: The New Science of Networks* (Barabási) | 2002 | PF,2,6 |
| *Evolution and structure of the Internet: A statistical physics approach* (Pastor-Satorras&Vespignani) | 2004 | 2,7 |
| *Atlas of Science: Visualizing What We Know* (Börner) | 2010 | FW,PF,8 |

Figures 2-4 show only nine of the listed books as the others are not connected to the highly interlinked 19 subject headings. For example, the *Atlas of Science* (Börner, 2010) has subject headings: *Classification of sciences – Atlases*, *Science – Atlases*, *Communication in science – Data processing*, *Digital mapping*. Only *Science – xxx* occurs in the network shown in Figures 2-4. However, the subject *Science – Atlas* was not merged under the term Science in the initial subject heading aggregation process and hence the book does not show in the figures.

Importantly, subject heading often are complex. That means, subject headings indicate multiple aspects of the topics covered in a work. Complex subject headings may be used to combine multiple topical subjects into one or to combine topical subjects with a specific methodology; temporal period and era; the material, format, and genre; or geographic regions and languages of a work. Many subject heading schemes have a hierarchical structure, e.g., a subject term has parent and child terms, or is a composite of two parent terms with two facets of co-equal status. In other words, there are relationships between subjects, books and classification schemes that are currently not utilized in this initial analysis. A follow-up study could refine links between books, classification schemes, and domains by using both topical and methodological subject headings.

There are also problems connected with the context-richness of subject headings. One problem is the vagueness of compound subject terms. Subjects headings like "Data processing", "Methodology", and "Research" and can describe many ideas and techniques. On the other side, compound subjects can allow for the collection of thematically related materials for later comparison. More specific methodological subjects, like network analysis, may be used as to compare the use of a method or technique across disciplines.

While the bibliographic-bibliometric method proposed and exemplified here benefits from the collective wisdom of librarians worldwide it also comes with its own caveats. However, the resulting analyses and visualizations can be used to gain a new, more comprehensive understanding of the richness of scientific disciplines, methods, and perspectives as captured in books.

**Conclusions**

The edited book *Models of Science Dynamics* provided a review of major models of science for an expert audience (Scharnhorst et al., 2012). This chapter introduced and exemplified a novel means to construct the topical or concept space in which works on models of science are situated by using key books, library subject headings, and classification codes. Specifically, this chapter *extends* existing methods of bibliometric analysis of classification systems to subject headings, which come from multiple controlled vocabularies. We implemented a method to identify and classify both book LC classifications and subject headings that uses a common framework of scientific domains to facilitate comparative analysis of book classification and subject headings.

Our method as applied to subject headings is unique in that it reveals a degree of cross-domain pollination of concepts and method that is not captured with LCC numbers. While LCC numbers reveal the most uniquely domain associated with a work, for our book data is in general science, the social science, math and physics, our analysis of subject heading reveals the domains are likely to create models and concepts used by other researchers working in other domains.

The bibliographic-bibliometric analysis of existing models of science provides a first depiction of major *disciplines*, *methods*, and *perspectives*. The study also highlights challenges and opportunities that arise when books, cataloging data, and subject headings are used in delineating and mapping a domain. It is our hope that this study inspires future reviews, exemplifications, and discussions of models of science developed in different scientific disciplines. Future work might expand this bibliographic-bibliometric analysis beyond books, e.g., to journal publications, course contents, and/or encyclopedias. It might attempt to generate cross-walks between science, engineering, education, and other classification systems and taxonomies that define and organize different model types. Likely, challenges encountered in the work presented here will persist—document titles and author names are non-unique, the terminology used differs considerably among the different disciplines, and among librarian-catalogers.

This chapter also contributes to work in information retrieval and domain delineation by helping to answer questions such as: What are the most important books on modelling science and how are these books positioned in the interdisciplinary landscape of science? Starting with the references in one book on models of science (Scharnhorst et al. 2012), a landscape unfolds as diverse and broad as the table of contents in Bernal's book (Bernal 1939). However, a comparison of the headings of Bernal's book and the dominant subject headings in Figure 3 reveals an important difference. The structure of Bernal's book reads like a *what-to-be-modeled* list. Examples are organization (*The existing organization of research in Britain*), scientific practices (*The efficiency of scientific research*), scientific careers (*The training of the scientist*) or globalization (*International science*). In turn, the dominant subject headings form a checklist of necessary dimensions or ingredients for a *good* model of science, a *how-to-model* list. Such a model would need to address the epistemic foundations of science (*Science Philosophy*), its social structure (*Science Social*), its relations to innovations and economic growth (*Innovation*), and aspects of its networked nature (*Systems theory, Social Networks, Internet*). The network of books and subject headings presented in this chapter gives a first orientation what to read with respect to this checklist. Ultimately, close reading, further tracing references, and original research will be needed to advance existing models of science.

## Acknowledgments

We would like to thank Cassidy Sugimoto for inspiring and editing this Festschrift and for her editorial comments. We thank David Kloster for his help collecting and coding subject heading data. Allyson Carlyle, Alexander Petersen, Richard Smiraglia, and Nicolai Vitanov provided expert comments on an earlier version of this chapter. The Sci2 Tool used in this study was developed by Chin Hua Kong, Adam Simpson, Steven Corenflos, Joseph Biberstine, Thomas G. Smith, David M. Coe, Micah W. Linnemeier, Patrick A. Phillips, Chintan Tank, and Russell J. Duhon. The Sci2 Tool uses the Cyberinfrastructure Shell (http://cishell.org) developed at the Cyberinfrastructure for Network Science Center (http://cns.iu.edu) at Indiana University. This work was partially funded by the National Institutes of Health under awards NIA P01AG039347 and U01 GM098959; and partly funded by the COST Action TD1210 KnowEscape.

## Appendix A

Bourdieu, P. (1986). Forms of capital. In: Richardson JG (Ed) *Handbook of theory and research for the sociology of education*. Greenwood, New York, NY, pp 241-258.

Brauer, F., Castillo-Chavez, C. (2001). *Mathematical models in population biology and epidemiology*. Texts in Applied Mathematics, Vol. 40. Springer, New York, NY.

Bryman, A. (1988). *Quantity and quality in social research*. Contemporary Social Research Series, Vol. 18. Unwin Hyman, London.

Bucchi, M. (2004). *Science in society: An introduction to social studies of science*. Routledge, London. Revised and expanded ed. of: Bucchi, M. (2002). *Scienza e Società: Introduzione alla sociologia della scienza*. Il Mulino, Bologna.

Budd, T. (2003). *Alcohol related assault*. London: Home Office.

C. T. J. Flood-Page (Ed.). (2003). *Crime in England and Wales 2001/2002: Supplementary Volume*. Home Office Statistical Bulletin 01/03. London: Home Office.

Caldarelli, G. (2007). *Scale-free networks: Complex webs in nature and technology*. Oxford Finance Series. Oxford University Press, Oxford.

Camerer, C. (2003). *Behavioral game theory experiments in strategic interaction*. New York [u.a.]: Russell Sage [u.a.].

Campbell, D.T. (1974). Evolutionary epistemology. In: Schilpp, P.A. (Ed.) *The philosophy of Karl Popper*. The Library of Living Philosophers, Vol. 14. Open Court Publishing, La Salle, IL.

Carrington, P. J., Scott, J., & Wasserman, S. (2005). Models and Methods in Social Network Analysis. Cambridge University Press.

Carruthers, P., Stich, S., Siegal, M. (Eds.) (2002). *The cognitive basis of science*. Cambridge University Press, Cambridge, pp 285-299 (DOI: 10.1017/CBO9780511613517.016).

Chapman, R.N. (1931). *Animal ecology*. McGraw Hill, New York, NY, pp 409-448.

Cole, S. & Cole, J.R. (1973). *Social stratification in science*. University of Chicago Press, Chicago, IL.

Coleman, J.S. (1964). *An introduction to mathematical sociology*. Free Press, New York, NY; Collier-Macmillan, London.

Youniss, J. & Smollar, J. (1985). *Adolescent relations with mothers, fathers, and friends*. Chicago: University of Chicago Press.

Ziman, J. (1994). *Prometheus bound: Science in dynamic steady state*. Cambridge University Press, Cambridge.

Ziman, J. (2000). *Real science: What it is, and what it means*. Cambridge University Press, Cambridge (DOI: 10.1017/CBO9780511541391).

Zipf, G.K. (1949). *Human behavior and the principle of least effort: An introduction to human ecology*. Addison-Wesley Press, Cambridge, MA.
38